\documentclass[aps, prb, reprint]{revtex4-2} 
\usepackage{fix-cm}
\usepackage[T1]{fontenc}

\usepackage{amsmath}  
\usepackage{amsfonts} 
\usepackage{graphicx} 
\usepackage{dcolumn}
\usepackage{bm}
\usepackage{xcolor}

\usepackage{titlesec}
\titlespacing*{\section}{0pt}{\baselineskip}{\baselineskip}
\usepackage[utf8]{inputenc}
\usepackage[T1]{fontenc}
\usepackage{mathptmx}
\usepackage{etoolbox}

\begin{document}


\title{The asymmetric rotating saddle potential as a 
mechanical analog to the RF Paul trap}


\author{Aidan Carey}
\email{aidanbcarey@gmail.com}
\author{Laurel Barnett}
\email{ljbarnett@college.harvard.edu}
\author{Robert Hart}
\email{roberthart56@gmail.com}
\author{Anna Klales}
\email{aklales@g.harvard.edu}
\author{Ali Kurmus}
\email{akurmus@uw.edu}
\author{Louis Deslauriers}
\email{louis@physics.harvard.edu}
\affiliation{Department of Physics, Harvard University, Cambridge, MA 02138}


\date{\today}

\begin{abstract}

Under specific conditions, a rotating saddle potential can confine the motion of a particle on its surface. This time-varying hyperbolic potential shares key characteristics with the RF-electric-quadrupole ion trap (RF Paul trap), making it a valuable mechanical analog. Previous work has primarily focused on symmetric saddles, characterized by equal curvatures along the trapping and anti-trapping directions. However, most applications of RF Paul traps—such as atomic clocks, quantum computing, and quantum simulations—require asymmetry in the quadrupole potential to break the degeneracy of motional modes, which is essential for processes like laser cooling and other quantum manipulations. In this paper, we investigate the motion of trapped particles in asymmetric rotating saddles. We demonstrate that even minor asymmetries, including those arising from manufacturing imperfections, can significantly affect particle trajectories and stability. Our analysis includes both theoretical modeling and experimental measurements. We derive the equations of motion for asymmetric saddles and solve them to explore stability and precession effects. Additionally, we present lifetime measurements of particles in saddles with varying degrees of asymmetry to map key features of the \( a \)-\( q \) stability diagram, including counterintuitive demonstrations of stability for saddles with negative asymmetry. This study underscores the importance of incorporating asymmetry into mechanical models of ion traps to better reflect real-world implementations. Although motivated primarily by RF Paul traps, these asymmetry-related results are also relevant to emerging gravitational analogs, such as rotating saddle potentials in certain binary black hole systems.

\end{abstract}

\maketitle 

\vspace{-10pt}
\section{INTRODUCTION}\label{sec:intro} 
 
A saddle-shaped surface rotating about its inflection point can confine a particle near its center. This otherwise unstable mechanical system becomes stable when the saddle surface rotates with an angular velocity that is tuned precisely to maintain the stability of the particle’s trajectory \citep{Paul_1990, Rueckner_1995}. Such dynamic stabilization is a central feature of many physical systems in both classical and modern physics, such as the linear stability of the triangular Lagrange equilibrium points (L4 and L5), the confinement of electrons in nonspreading Trojans—Rydberg wave packets stabilized by a rotating electric field, and the confinement of charged particles in rapidly oscillating quadrupole electric fields, such as those found in RF-electric-quadrupole ion traps (RF Paul traps). These represent just a few of the many systems in nature that rely on dynamic stabilization mechanisms \cite{Kapitza_1951, Whymark_1975, Raab_1987, Bialynicki-Birula_1994, Murray_1999, Dehmelt_1967, Leibfried_2003, Abe_2010}. 

The rotating saddle trap was first identified as a mechanical analog to RF Paul traps in 1989 by Wolfgang Paul during his Nobel lecture \cite{Paul_1990}. RF Paul traps are powerful tools for studying quantum systems, utilizing time-varying hyperbolic potentials to dynamically stabilize and trap particles \cite{Wineland_1998}. Although there are differences in how these potentials vary in time, the rotating saddle trap serves as an invaluable tool for understanding the principles of the RF Paul trap.

Recently, it has been shown that the effective gravitational potential near the center of a charged binary black hole system (cBBH) forms a dynamically rotating saddle, mathematically identical to the mechanical rotating saddle studied in laboratory experiments. This exact correspondence suggests that insights gained from mechanical rotating saddles may have direct implications for plasma confinement in astrophysical settings \citep{Kurmus_2025}. 

Previous research has focused on symmetric saddles, characterized by equal curvatures along the trapping and anti-trapping directions \cite{Paul_1990, Rueckner_1995, Thompson_2002, Kirillov_2016, Fan_2017, Borisov_2018, Lofgren_2023}. However, most applications using RF Paul traps, such as atomic clocks and quantum computing \cite{Chou_2010, Ladd_2010}, require asymmetry in the saddle-like electric potential. This asymmetry is necessary for breaking the degeneracy of motional modes, which is crucial for processes like laser cooling and various quantum manipulations \cite{Deslauriers_2004, Leibfried_2003}. 

Moreover, asymmetry is not only an engineered feature of quantum traps but also an intrinsic property of the gravitational saddle potential in a charged binary black hole system \citep{Kurmus_2025}. Gauss's law dictates that the curvature of the effective potential must differ along the three spatial directions, meaning that even in an idealized case, the rotating gravitational saddle in a cBBH system is fundamentally asymmetric. 

Furthermore, even small manufacturing defects in physical rotating saddles—on the order of tens of microns—can introduce such asymmetries that significantly alter particle trajectories and stability. As we will show, such imperfections can have a significant impact, leading to deviations in motion that are difficult to predict without explicitly accounting for asymmetry. Therefore, considering asymmetry in rotating saddles is crucial—not only because true analogs to RF Paul traps must be asymmetric, but also because even small deviations can result in substantial changes in particle behavior.

In this paper, we systematically examine how asymmetry in the hyperbolic shape of a rotating saddle impacts particle dynamics. Using lifetime measurements of a steel ball-bearing in saddles with varying degrees of asymmetry, we map key features of the stability diagram and characterize how asymmetry alters confinement and precession. This study underscores the necessity of accounting for asymmetry—not only to better reflect real-world RF Paul traps and the inherently asymmetric potentials in binary black hole systems, but also because even small asymmetries can lead to qualitatively different trajectories.

\section{COMPARING THE RF PAUL TRAP AND THE ROTATING SADDLE }\label{sec:rf_vs_rotating}

\begin{figure*}[!t]
    \centering
    \includegraphics[width=0.70\textwidth, angle=0,]{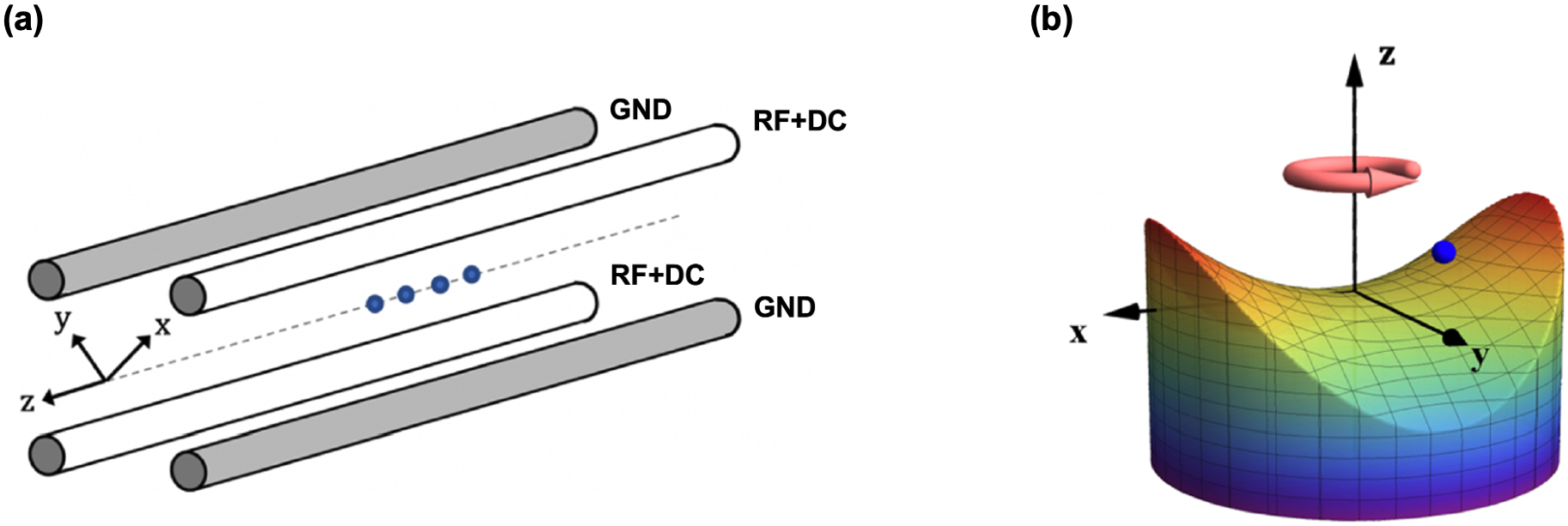}
    \caption{(a) Schematic of a two-dimensional RF Paul trap, showing the electrode configuration that creates a saddle-shaped, time-varying potential in the \( x,y \) plane. A positive ion experiences alternating trapping and anti-trapping along the \( x \)- and \( y \)-directions, respectively, due to the “flapping potential.” The same electrodes that generate the RF signal (\( U_0 \cos \Omega t \)) are also biased at a DC voltage \( U_{\text{DC}} \), causing the trap rods to “float” at a nonzero potential. This breaks the symmetry of the quadrupole potential and introduces a static curvature along each axis. (b) Illustration of a rotating saddle surface with upward curvature (trapping) along the \( x \)-axis and downward curvature (anti-trapping) along the \( y \)-axis, characterized by saddle radius \( r_0 \) and height \( h_0 \) along the trapping direction.}
    \label{fig:4_rods_and_saddle}
\end{figure*} 

To better understand the importance of asymmetry in rotating saddles, it helps to first examine how asymmetry arises in RF Paul traps and its impact on particle motion and stability. These devices are intentionally designed with asymmetry to lift degeneracies in motional modes, which refer to the distinct vibrational modes of particles around their equilibrium positions within the trap. Breaking these degeneracies is crucial for enabling applications such as laser cooling and precise quantum manipulations \cite{Wineland_1998, Deslauriers_2004}. By examining the origins and effects of asymmetry in RF Paul traps, we can better appreciate the parallel significance of asymmetry in rotating saddle systems.

The RF Paul trap utilizes a time-varying quadrupole electric potential generated by electrodes arranged as illustrated in Fig.~\ref{fig:4_rods_and_saddle}(a). According to Earnshaw’s theorem, stable equilibrium for a charged particle cannot be achieved using static electric fields alone. However, by employing oscillating electric fields that alternate between trapping and anti-trapping along perpendicular axes in the $xy$-plane, the RF Paul trap achieves dynamic stability. This alternating, or “flapping,” potential effectively produces a time-averaged confining force, stabilizing the particle’s motion \cite{Paul_1990, Thompson_2002}.

This comparison highlights how the intentional use of asymmetry in RF Paul traps motivates a parallel exploration in rotating saddles, setting the stage for a detailed analysis of how asymmetry affects stability and confinement in the latter system.


\subsection{RF Paul Trap Equations}

The RF Paul trap’s potential, illustrated schematically in Fig.~\ref{fig:4_rods_and_saddle}(a), is given mathematically by a time-dependent quadrupole potential:

\begin{equation}
\label{eq:RFPotential}
U(x, y, t) = \frac{\left( U_{\text{DC}} + U_0 \cos \Omega t \right)}{r_0^2} \left( x^2 - y^2 \right). 
\end{equation}

Applying Newton’s second law to a charged particle (mass $m$, charge $e$) in this potential yields the Mathieu equations \cite{Wineland_1998} 

\begin{subequations}
\label{eq:mathieu}
\begin{align}
\frac{d^2 x}{d \tau^2} + 2q \cos(2 \tau) x + a x &= 0, \label{eq:mathieu_x}\\
\frac{d^2 y}{d \tau^2} - 2q \cos(2 \tau) y - a y &= 0. \label{eq:mathieu_y}
\end{align}
\end{subequations}

where \( \tau = \frac{\Omega t}{2} \), \( q = \frac{4 e U_0}{m r_0^2 \Omega^2} \), and \( a = \frac{8 e U_{\text{DC}}}{m r_0^2 \Omega^2} \).

The parameters \( q \) and \( a \) in Eqs. \eqref{eq:mathieu_x} and \eqref{eq:mathieu_y} represent the "flapping" and "static" contributions to the potential, respectively. Parameter \( q \) specifies the magnitude of the trap's oscillatory force, while \( a \) determines the direction and magnitude of the time-invariant force acting on the ion.

\subsection{Rotating Saddle Equations}

In the rotating saddle configuration (Fig.~\ref{fig:4_rods_and_saddle}(b)), the gravitational potential in the rotating reference frame \( (x', y') \) is given by

\begin{equation}
\label{eq:SaddlePotential}
U(x', y', t = 0) = \frac{mg h_0}{r_0^2} \left( \beta x'^2 - y'^2 \right). 
\end{equation}

Here, \( \beta \) is a dimensionless parameter representing the saddle's asymmetry, while \( h_0 \) and \( r_0 \) are geometric parameters describing its curvature. Assuming a symmetric saddle with \( \beta = 1 \), we interpret \( h_0 \) and \( r_0 \) as follows: at a distance \( r_0 \) along the \( x \)-axis, the saddle reaches a height \( h_0 \). Transforming this into the lab frame rotating at frequency \( \Omega \), the equations become:

\begin{subequations}
\label{eq:SaddleMathieu}
\begin{align}
\frac{d^2 x}{d t^2} + \left( a + 2q \cos(2t) \right) x + 2q \sin(2t) y &= 0, \label{eq:saddle_mathieu_x}\\
\frac{d^2 y}{d t^2} + \left( a - 2q \cos(2t) \right) y + 2q \sin(2t) x &= 0. \label{eq:saddle_mathieu_y}
\end{align}
\end{subequations}

with the substitutions \( \tau \equiv \Omega t \), \( a = \frac{K(\beta - 1)}{m\Omega^2} \), \( q = \frac{K (\beta + 1)}{2  m\Omega^2} \), and \( K = \frac{m g h_0}{r_0^2} \).

Comparing Eqs. \eqref{eq:mathieu_x} and \eqref{eq:mathieu_y} with Eqs. \eqref{eq:saddle_mathieu_x} and \eqref{eq:saddle_mathieu_y}, both sets of equations share structural similarities. Each includes sinusoidal terms proportional to \( q \) and a time-invariant term proportional to \( a \). However, unlike Eqs. \eqref{eq:mathieu_x} and \eqref{eq:mathieu_y}, equations Eqs. \eqref{eq:saddle_mathieu_x} and \eqref{eq:saddle_mathieu_y} are coupled.

\section{COMPLEX REPRESENTATION AND STABILITY CONDITIONS OF THE ROTATING SADDLE}

To analyze particle dynamics in the rotating saddle potential, we introduce a complex representation that simplifies the system of differential equations \cite{Thompson_2002}. Defining the complex variable \( z = x + i y \), the equations of motion \eqref{eq:saddle_mathieu_x} and \eqref{eq:saddle_mathieu_y} can be reformulated into a single complex differential equation:

\begin{equation}
\label{eq:ComplexEqtnMotion}
\frac{d^2 z}{d \tau^2} + az + 2qz^* e^{i 2 \tau} = 0. 
\end{equation}

The appearance of both \( z \) and its complex conjugate \( z^* \) in equation \eqref{eq:ComplexEqtnMotion} reflects the coupling between the \( x \) and \( y \) components in the original system. To solve this equation, we first take the complex conjugate of (5), which introduces terms involving \(z^{\prime *}\) and \(\quad z^{\prime\prime *}\)). These are eliminated by expressing \( z^* \) in terms of \( z \) using the original equation and differentiating as needed, leading to a fourth-order differential equation. Applying the ansatz \( z(\tau) = e^{i \tau} f(\tau) \), which captures the slower varying motion in a rotating frame, simplifies the solution process. The general solution is then given by:

\begin{equation}
\label{eq:ComplexSoln}
z(\tau) = e^{i \tau} \left[ A e^{i \alpha_- \tau} + B e^{-i \alpha_- \tau} + C e^{i \alpha_+ \tau} + D e^{-i \alpha_+ \tau} \right] 
\end{equation}

where $A$, $B$, $C$, and $D$ are complex constants determined by initial conditions. To determine these integration constants, we substitute the initial conditions for position and velocity into Eq.~\eqref{eq:ComplexSoln}, yielding a system of four equations from which the coefficients can be uniquely determined. The characteristic exponents $\alpha_\pm$ are given by: 

\begin{equation}
\label{eq:Exponents}
\alpha_\pm = \sqrt{(1 + a) \pm 2 \sqrt{a + q^2}}. 
\end{equation}

\begin{figure*}[!t]
    \centering
    \includegraphics[width=0.65\textwidth, angle=0,]{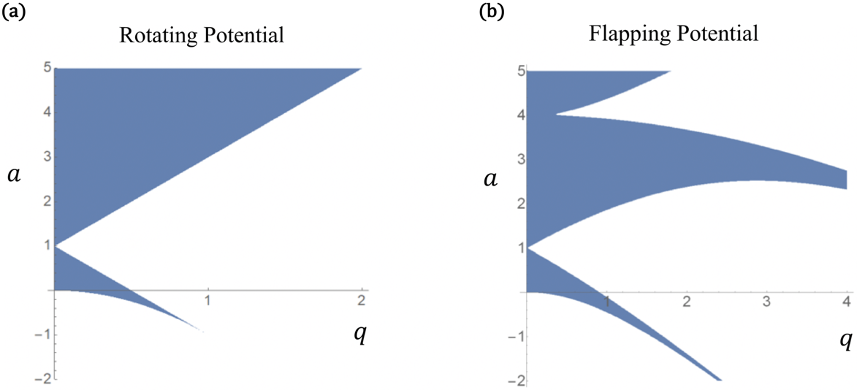}
    \caption{(a) Stability diagrams for (a) the rotating saddle potential and (b) the RF Paul trap. Blue shaded regions indicate parameter combinations of $a$ and $q$ that yield stable particle trajectories. The rotating saddle's analytically solvable equations produce a simpler triangular stability region for positive $a$ with instability only at $a=1$, while the RF Paul trap's non-analytical Mathieu equations result in a more complex stability boundary with multiple instability regions. At negative $a$ values, the rotating saddle stability terminates at $q=1$,$a=-1$ whereas the RF Paul trap's stability region extends further.}
    \label{fig:Figure_aq_diagram_RF_and_Saddle}
\end{figure*} 

\begin{figure*}
    \centering
    \includegraphics[width=0.80\textwidth, angle=0,]{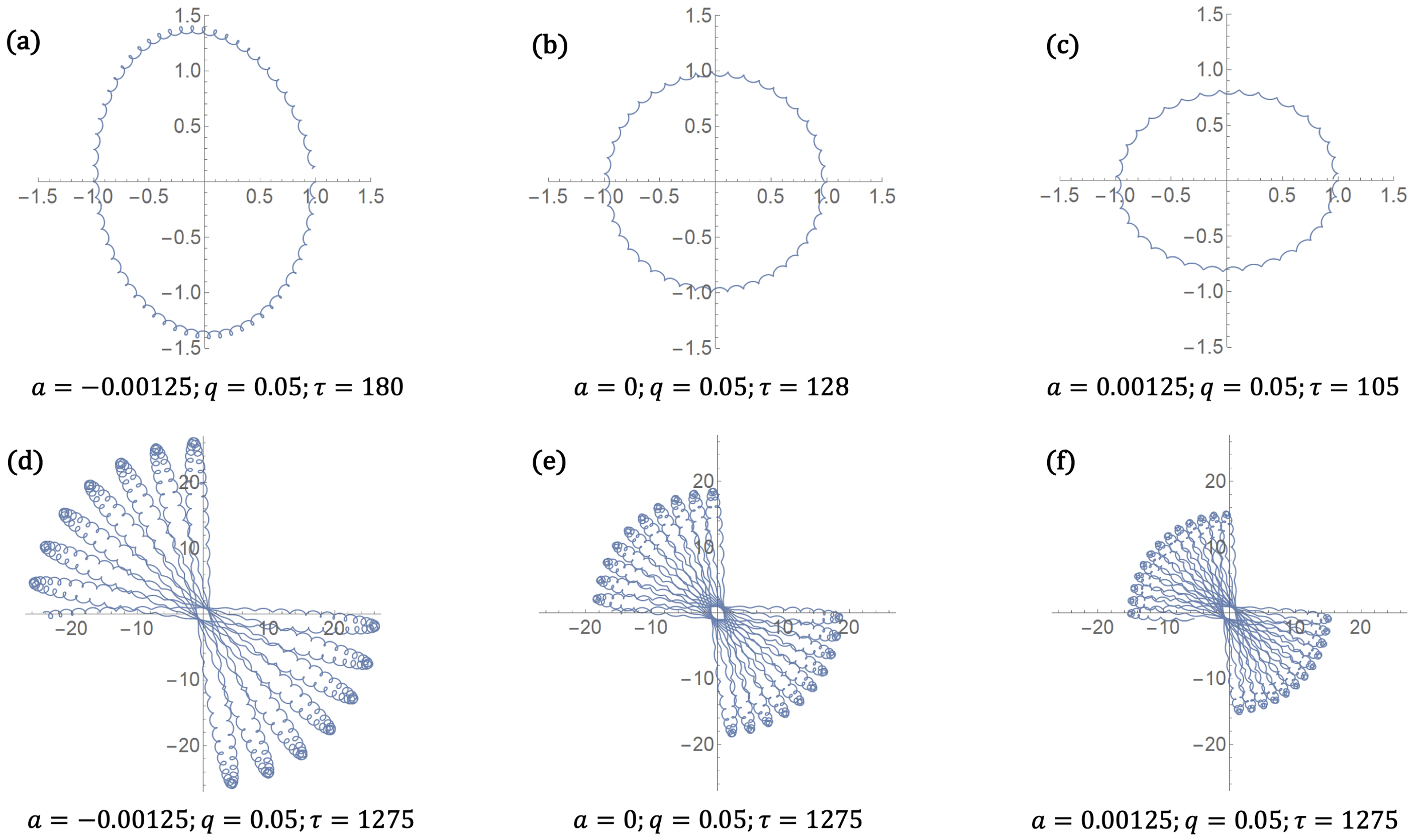}
    \caption{Particle trajectories for different initial conditions $(z_0, v_0)$ and values of the asymmetry parameter $a$. Top row: $z(0) = 1$, $z'(0) = 0$. Bottom row: $z(0) = 1$, $z'(0) = i$. The asymmetry parameter \( a \) varies across columns, while \( q \) is fixed at 0.05. The trajectory duration $\tau$ is indicated on each panel. Minor variations in asymmetry parameter $a$ are seen to cause significant alterations to particle motion}.
    \label{fig:trajectory_diagram}
\end{figure*}

The stability of particle trajectories is determined by the nature of the characteristic exponents \( \alpha_\pm \). When \( \alpha_\pm \) is complex, the motion exhibits an exponentially growing contribution, indicating instability. Conversely, if \( \alpha_\pm \) is real, the motion remains bounded and stable. The shaded regions in Fig.~\ref{fig:Figure_aq_diagram_RF_and_Saddle} illustrate these stability conditions: panel (a) shows the stable parameter space for the rotating saddle, while panel (b) presents the analogous stability regions for the RF Paul trap. Although both systems share qualitative similarities, their stability regions differ in shape and extent, reflecting fundamental differences in their underlying dynamics and the coupling present in the rotating saddle equations. Notably, for \( a = 1 \), both traps exhibit a parametric resonance that prohibits stable orbits for all values of $q$, though the RF Paul trap additionally shows resonant instabilities at higher integer values of $a$, and exhibits curved rather than linear boundaries in its stability regions due to the more complex, non-analytic nature of its solutions. A similar trend emerges for negative values of $a$, where stability regions extend as $q$ increases and $a$ decreases. However, this behavior diverges between the two systems: stability in the rotating saddle is lost at \( q = 1, a = -1 \), whereas in the flapping potential, stability extends indefinitely—albeit within a rapidly shrinking parameter space at increasingly large values of \( q \) and negative \( a \). This lower half of the stability diagram is particularly interesting: although a negative \( a \) corresponds to a time-averaged potential that is unstable (i.e., an inverted parabola), stability is still achieved when the ponderomotive effect from \( q \) becomes sufficiently strong to overcome this inherent instability. This explains the distinctive shape of the shaded region for \( a < 0 \), where increasing \( q \) is required to compensate for increasingly negative \( a \). Later in this paper, we experimentally confirm this behavior by demonstrating particle confinement in rotating saddles with negative \( a \).

\subsection{Impact of Asymmetry on Particle Dynamics}

The asymmetry parameter \( a \) plays a crucial role in determining the behavior of particles within the rotating saddle potential. This parameter quantifies the imbalance between the positive and negative curvatures of the saddle surface, and even small deviations from perfect symmetry can significantly alter particle trajectories.

Figure~\ref{fig:trajectory_diagram} shows simulated particle trajectories for three values of the asymmetry parameter $a$, illustrating how even small deviations from symmetry affect motion. The top-row trajectories exhibit nearly circular motion, which differs from the more commonly presented flower-like trajectories shown in the bottom row. This contrast arises from the initial conditions: when the particle starts at rest at \( z(0) = 1 \), the initial force is directed radially inward (see Eq. (5)). However, because the driving term \( 2qz^*e^{i2\tau} \) rotates in time, the force vector sweeps ahead of the particle's response. As a result, the particle begins to lag behind the rotating force, and its trajectory becomes azimuthal, leading to a stable swirling motion at roughly constant radius. This behavior reflects a fundamental distinction between rotating saddles and RF Paul traps: in the rotating saddle, the time-dependent force rotates and couples the \( x \)- and \( y \)-directions, whereas in RF Paul traps the \( x \)- and \( y \)-components are driven independently. Although the top-row trajectories appear nearly circular, close inspection reveals they do not exactly close—precession is still present, albeit more subtly than in the bottom-row trajectories where initial velocity makes it more visually apparent.

This dependence on initial conditions aside, the excursion of the particle away from the origin is strongly influenced by the asymmetry parameter \( a \). For \( a = 0 \), the time-averaged ponderomotive potential provides the sole restoring force, resulting in trajectories that remain bounded and relatively compact. When \( a > 0 \), the addition of a static inward potential enhances confinement, effectively steepening the potential well and leading to tighter orbits with reduced radial excursion. This effect is evident in both the zero-velocity (top row) and finite-velocity (bottom row) cases.

Conversely, for \( a < 0 \), the static component of the potential becomes destabilizing, acting outward and opposing the ponderomotive confinement. This reduces the net restoring force and allows the particle to explore a larger region of phase space. As shown in panels (a) and (d), the trajectories become significantly more extended, and in some cases, only marginally bounded.

\subsection{The necessity of considering asymmetry}

The substantial variation in trajectories observed in Fig.~\ref{fig:trajectory_diagram} for values of \( a \) as small as \( \pm 0.00125 \) highlights the extreme sensitivity of particle motion to asymmetry in the saddle’s curvature. This is particularly striking given that \( a = 0.00125 \) corresponds to less than a 0.1\% difference in curvature between the trapping and anti-trapping directions. For a typical experimental setup with a saddle radius \( r_0 = 15 \) cm, height \( h_0 = 2.5 \) cm, and rotation frequency \( \Omega \sim 1 \) Hz, this level of asymmetry translates to a mere 25-micron difference in saddle height along orthogonal axes—an order of magnitude smaller than the width of a human hair. Despite the minute nature of this deviation, the resulting changes in trajectory confinement and precession are substantial, emphasizing that even minuscule deviations—well below visual resolution—can noticeably affect the resulting dynamics.

In fact, this sensitivity becomes even more pronounced in certain regions of the stability diagram. For example, with \( q = 0.05 \), we find that reducing the asymmetry parameter from \( a = -0.00125 \)—as shown in Fig.~\ref{fig:trajectory_diagram}—to \( a = -0.00185 \) results in a doubling of the particle's maximum excursion from the origin. A further decrease to \( a = -0.00250 \) leads to complete instability and ejection from the saddle. While this region of parameter space is already known to be sensitive, these results highlight just how abruptly stability can degrade as asymmetry increases. For this reason, we recommend building physical saddles with a slight positive bias in \( a \), which ensures the system remains safely within the stable region even if minor manufacturing imperfections reduce the curvature asymmetry. This approach mirrors the intentional asymmetry found in RF Paul traps and offers a practical design strategy for maintaining stability in the presence of unavoidable imperfections.

Unlike the RF Paul trap, where imperfections in electrode geometry result in relatively minor perturbations to the trapping potential due to the electrodes being positioned farther from the trap center, deviations in the physical surface of a rotating saddle directly modify the effective potential landscape at the location where the particle is confined. Any deviation from perfect symmetry—whether due to machining tolerances or possibly thermal expansion—modifies the particle’s motion in ways that cannot be ignored in experiments designed for higher precision. This intrinsic sensitivity highlights the limitations of models that assume perfect symmetry, particularly when used to interpret experimental data.

However, the most likely manifestation of asymmetry in a physical saddle is not a sharp or random deviation in curvature, but a smooth large-scale distortion introduced during fabrication. In practice, these surfaces are typically 3D printed, coated, and then hand-sanded to reduce roughness. While this process suppresses high-frequency surface imperfections, it can introduce gradual warping or curvature variation over several centimeters. These low-spatial-frequency deviations are difficult to characterize precisely, but they often resemble the smoothly asymmetric saddles modeled in this work. Importantly, because the ball rarely moves more than a few centimeters from the trap center, it effectively samples a local surface that appears smooth, even if the global curvature deviates from perfect symmetry. Given these considerations, our choice to model asymmetry as a smooth, controlled deviation is not only analytically tractable but also representative of the dominant imperfections likely to affect particle motion in real-world experiments.

\subsection{Secular Motion and Time-Averaged Dynamics}

In the limit of small \( a \) and \( q \), the motion of a particle in the rotating saddle naturally decomposes into a rapidly oscillating micromotion superimposed on a slower secular motion. This terminology, borrowed from RF Paul traps \cite{Wineland_1998}, provides a useful framework for analyzing long-term behavior by averaging over micromotion. The secular motion, which governs the overall confinement of the particle, exhibits features characteristic of a two-dimensional harmonic oscillator, but is modified by the asymmetry parameter \( a \).

As seen in Equation (5), the general solution for \( z(\tau) \) consists of four oscillatory terms. In the small \( a, q \) limit, where \( \alpha_\pm \approx 1 \), multiplying by \( e^{i\tau} \) results in two fast and two slow components. Averaging over micromotion removes the fast terms, leaving the time-averaged secular motion:
\begin{equation}
z_s(\tau)= B e^{i(1-\alpha_- )\tau} + D e^{i(1-\alpha_+) \tau}.
\end{equation}

A key distinction between the rotating saddle and its RF Paul trap analog is the presence of precession in the rotating saddle system \cite{Kirillov_2016}. Unlike Paul traps, where stable trajectories remain fixed in orientation, trajectories in the rotating saddle exhibit a slow precessional motion in the lab frame. This precession, a fundamental feature of rotating saddles, arises due to an effective Coriolis-like force induced by the potential’s rotation \cite{Kirillov_2017}, and it has been shown to be mathematically analogous to the precession observed in a Foucault pendulum \cite{Cox_2020, Kirillov_2016}.

Noting the similarity to elliptical motion in the complex plane, we introduce the precession frequency \( \omega_p \) and the secular frequency \( \omega_s \):

\begin{equation}
z_s (\tau)= e^{i\omega_p \tau} \left( B e^{i\omega_s \tau} + D e^{-i\omega_s \tau} \right),
\end{equation}
where the precession and secular frequencies are given by

\begin{equation}
\omega_p = 1-\frac{\alpha_+ + \alpha_-}{2}, \quad \omega_s = \frac{\alpha_+ - \alpha_-}{2}.
\end{equation}

Since the secular motion is described by the sum of counter-rotating exponentials, the resulting trajectory forms an ellipse in the rotating frame. The additional phase factor \( e^{i\omega_p \tau} \) induces a slow precession of this ellipse in the lab frame.

Noting that \( \alpha_+ \) and \( \alpha_- \) are both functions of \( a \) and \( q \) (see Eq.~\eqref{eq:Exponents}), we expand \( \omega_s \) and \( \omega_p \) to second order:
\begin{equation}
\omega_s \approx \sqrt{a+q^2},
\end{equation}

\begin{equation}
\label{eq:OmegaPrecession}
\omega_p \approx \frac{1}{2} q^2+\frac{1}{8}a^2 .
\end{equation}

These expressions reveal that asymmetry not only modifies the stability of trajectories but also systematically shifts the characteristic frequencies of the system. The secular frequency \( \omega_s = \sqrt{a + q^2} \) increases monotonically with \( a \), meaning that even modest asymmetries can alter the effective trapping strength. 

By contrast, the precession frequency \( \omega_p \approx \frac{1}{2}q^2 + \frac{1}{8}a^2 \) depends quadratically on \( a \), making it less sensitive to small asymmetries than the secular frequency. Nonetheless, the effect is still present and can serve as a useful diagnostic for identifying deviations from perfect symmetry. In the limit of small \( a \) and \( q \), the secular motion closely resembles that of a two-dimensional harmonic oscillator, though with a slow precession superimposed on the elliptical motion. The corresponding effective trapping frequency is given by
\begin{equation}
\frac{k}{m} = 2\Omega \left( a + \frac{q^2}{2} \right).
\end{equation}

\begin{figure*}
    \centering
    \includegraphics[width=0.80\textwidth, angle=0,]{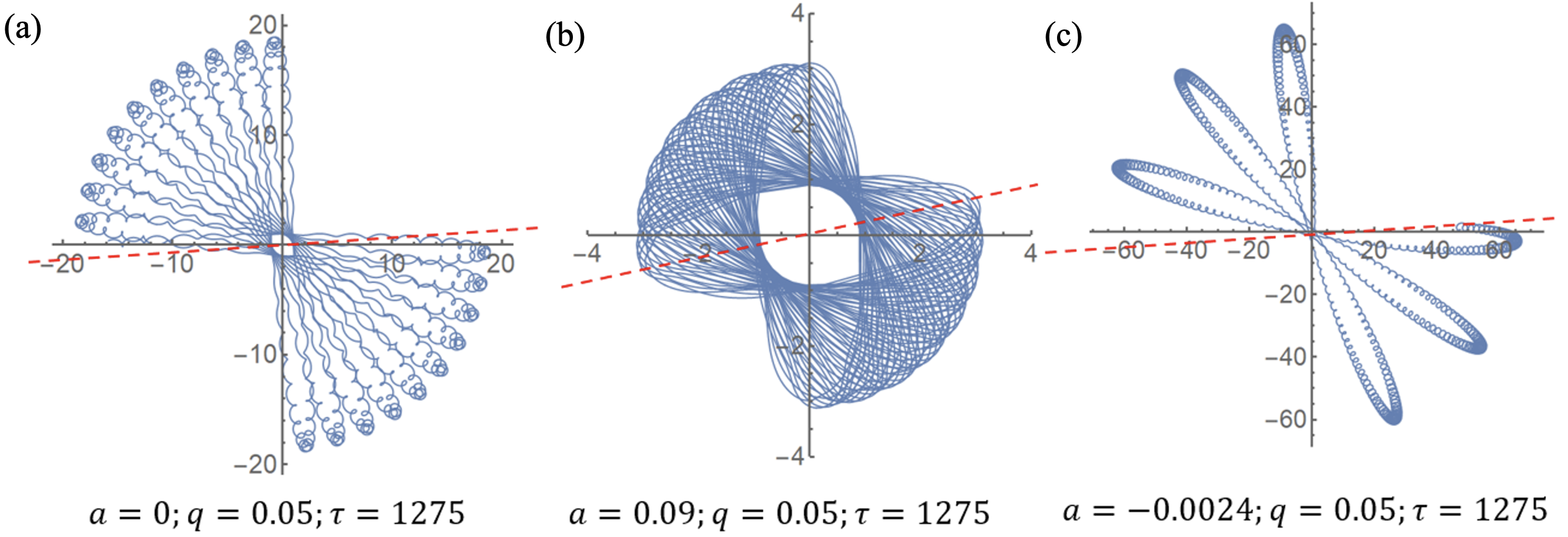}
    \caption{Particle trajectories illustrating the effect of asymmetry parameter \( a \) on precession. Each trajectory is evolved over the same time duration \( \tau = 1275 \) with a fixed parameter \( q = 0.05 \), placing the system deep in the secular regime. The initial conditions for the three trajectories are \( z(0) = 1, z'(0) = i \). The asymmetry parameter is varied across panels: (a) \( a = 0 \), (b) \( a = 0.09 \), and (c) \( a = -0.0024 \). The red line serves as a visual guide to indicate the precession angle \( \theta_p = \omega_p \tau \), with precession frequency given by \( \omega_p = \frac{1}{2} q^2 (1 + a + a^2) \). The precession angles for (a) and (c) are approximately the same, whereas for (b), with a larger asymmetry, the precession angle is noticeably increased.}
    \label{fig:Precession_plot}
\end{figure*}

The effect of asymmetry on precession is illustrated in Fig.~4, which shows particle trajectories evolved over the same time duration for different values of \( a \). The red line serves as a visual guide to indicate the precession angle \( \theta_p = \omega_p \tau \), where the precession frequency is given by Eq.~\eqref{eq:OmegaPrecession}. 

From the trajectories in Fig.~4, it is evident that for small \( a \), the precession frequency remains nearly unchanged, and the primary effect of asymmetry is seen in the secular frequency and range of motion. However, for larger values of \( a \) (Fig.~4b), the precession angle noticeably increases, demonstrating the quadratic dependence of \( \omega_p \) on \( a \). This effect is less pronounced for smaller asymmetries, as seen in Fig.~4c, where the precession angle remains close to that of the symmetric case in Fig.~4a.

\section{EXPERIMENT}

Previous experimental studies of the rotating saddle have been limited to the symmetric case, where stability was measured along the \( (a, q) = (0, q) \) line, varying \( q \) from small values up to near the instability threshold at \( q =0.5 \). In contrast, to systematically investigate the role of asymmetry in modifying stability conditions, we constructed saddles with different values of the asymmetry parameter \( \beta \). This allowed us to achieve nonzero values of \( a \) by varying the rotation frequency \( \Omega \). By measuring stability as a function of both \( a \) and \( q \), we mapped out stability regions beyond the symmetric case, providing a more complete picture of the system’s dynamics.

\subsection{Experimental Setup}

\begin{figure*}
    \centering
    \includegraphics[width=0.95\textwidth, angle=0,]{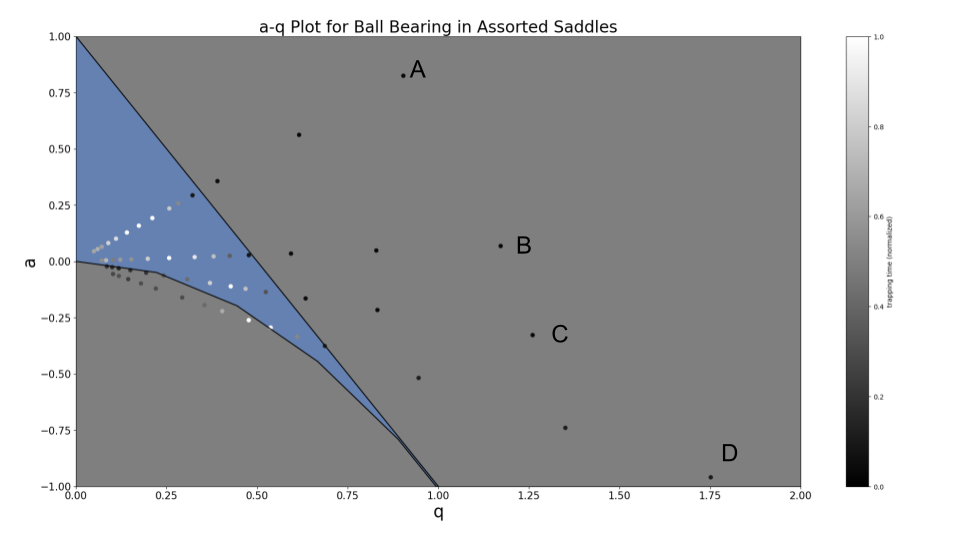}
    \caption{Normalized average particle lifetimes for four different physical saddles. Each radial line corresponds to a different saddle. The maximum lifetime measurement for each saddle is used to normalize the other data points in the series, so the figure shows \(\frac{t_{i,\beta}}{t_{max,\beta}}\)}.
    \label{fig:saddle_lifetime_plot}
\end{figure*}

To test the applicability of the stability regions in  Fig.~\ref{fig:Figure_aq_diagram_RF_and_Saddle}(a) to a real-world system, we constructed four physical saddles, each with a different asymmetry value \(\beta\). Each saddle was 3D-printed on an 18 cm square base, sanded, and coated with epoxy to minimize surface irregularities and improve measurement consistency. The saddles were then affixed to an axle driven by a stepper motor, enabling precise control of the rotational frequency.

With the trap rotating at a fixed frequency \( \Omega \), a ball bearing was placed at the center of the saddle, and the time until the ball fell off the trap—a proxy for stability—was recorded. For a given saddle geometry (\(h_0, r_0, \beta_i\)), varying the rotation frequency traces out a trajectory in the \(a,q\) stability diagram along the line:

\begin{equation}
\label{eq:aVSqEqtn}
    a = 2\frac{\beta_i - 1}{(\beta_i + 1)} q
\end{equation}

allowing us to experimentally map stability boundaries. The four saddles used in our experiments and their corresponding asymmetry coefficients \(\beta_i\) are summarized in Table~\ref{table:saddle_coefficients}.

\begin{table}[h]
    \centering
    \begin{tabular}{c c}
        \hline
        Saddle & \hspace{0.5cm} Asymmetry Coefficient (\(\beta\)) \\  
        \hline
        A & 2.68 \\
        B & 1.06 \\
        C & 0.77 \\
        D & 0.57 \\
        \hline
    \end{tabular}
    \caption{Coefficients for Saddles.}
    \label{table:saddle_coefficients}
\end{table}

While rolling ball experiments have traditionally served as physical analogs of the rotating saddle potential \cite{Paul_1990, Thompson_2002, Fan_2017}, limitations exist in reproducing theoretical behavior. Stability trends align well with predictions, but finer trajectory details, such as precession, are not directly observable due to intrinsic differences between the experimental and theoretical models \cite{Thompson_2002,Kirillov_2016,Kirillov_2017,Fan_2017}.

\subsection{Experimental Results and Discussion}

The results of our lifetime measurements across different saddles and stability parameters \(a\) and \(q\) are presented in Fig.~\ref{fig:saddle_lifetime_plot}, with each data point representing a single experimental trial. The normalized lifetimes, indicated by the grayscale intensity of each point, reveal a clear dependence of stability on saddle asymmetry. Points clustered along radial lines correspond to measurements from the same physical saddle at different rotation frequencies, effectively tracing specific $a$-$q$ trajectories through the stability diagram. White-colored dots indicate the longest measured lifetimes (normalized to 1.0 for each saddle), while darker dots correspond to shorter lifetimes, with black indicating immediate ejection from the saddle. For example, if the longest recorded lifetime for saddle A is 50 seconds, it receives a normalized value of 1.0 (white), while a shorter lifetime of 23 seconds would be assigned an intermediate grayscale value.

Points clustered along radial lines correspond to measurements from the same physical saddle at different rotation frequencies $\Omega$, effectively tracing specific $a-q$ trajectories through the stability diagram, in accordance with Eq.~\eqref{eq:aVSqEqtn}. In our experiments, the longest measured lifetimes were on the order of 50 seconds, primarily occurring near the center of the theoretically predicted stability regions. This effect was particularly pronounced for saddles with positive asymmetry parameter \(\beta\), namely saddles A and B. 

The lifetime measurements show a strong dependence on the stability boundary predicted in theory. Specifically, the data reveal a sharp transition from long-lived trajectories (white) to immediate ejection (black) near the line:

\begin{equation}
    a = -2q + 1
\end{equation}

This confirms that trajectories within the stability region, to the left of this boundary, exhibit prolonged confinement, whereas trajectories beyond this line are entirely unstable. The agreement between experimental results and theoretical predictions demonstrates that real asymmetric saddles behave as expected when incorporating the asymmetry parameter \( a \).

An especially noteworthy feature of the data is observed in saddles C and D, which possess negative asymmetry coefficients (\(\beta < 1\)). Unlike symmetric saddles, where the time-averaged potential over one rotation period is a flat plane, these saddles produce an inverted parabolic potential when averaged over one period. Naively, this would suggest intrinsic instability at all points, since a static saddle with negative \( a \) should always be unstable. However, our results show that stability can still be achieved for sufficiently large values of \( q \), indicating that the ponderomotive force generated by the saddle’s rotation is strong enough to counteract the net unstable potential imposed by negative \( a \). This is in stark contrast to saddles A and B, where stability exists at lower values of \( q \). Instead, for saddles C and D, stability only emerges when \( q \) is large enough, confirming that the ponderomotive effect must dominate over the inherent static instability.

While these experimental results provide a compelling validation of the theoretical model, it is important to acknowledge the inherent limitations of using rolling ball bearings as experimental probes. In particular, friction and rolling motion introduce deviations from the idealized point-mass dynamics assumed in theoretical treatments \cite{Fan_2017}. This effect is most pronounced at low values of \( a \) and \( q \), where theory predicts stability, but experimental data show instability across all cases. These conditions correspond to higher rotation frequencies, where rolling motion and friction likely dominate, leading to premature ejection of the ball bearing. Such non-ideal effects may also explain why some data points near the stability boundary \( a = -2q + 1 \) appear unstable.

\section{CONCLUSION}

This study demonstrates the crucial role of asymmetry in rotating saddle traps, revealing how even minute imperfections—on the order of tens of microns for a 10 cm saddle—can significantly alter stability conditions. Our results emphasize that real-world saddles are inherently asymmetric and that asymmetry must be considered when using mechanical models to approximate RF Paul traps. Since RF Paul traps always break the degeneracy of motional modes \cite{Wineland_1998, Leibfried_2003}, a truly analogous mechanical system must also be asymmetric. Moreover, recent work has shown that the effective gravitational potential in charged binary black hole systems forms an inherently asymmetric rotating saddle \citep{Kurmus_2025}, highlighting the astrophysical relevance of the asymmetries explored in this study.

Through lifetime measurements of ball bearings in saddles with varying asymmetry, we traced out the \( a,q \) stability diagram, confirming the strong influence of \( a \) on particle motion. The striking stability observed in saddles with negative \( a \) further highlights the role of the ponderomotive force in counteracting static instability. For cases where long-lived stability is the goal—such as classroom demonstrations—we suggest intentionally incorporating a slight positive asymmetry to guard against inadvertent imperfections that might otherwise introduce destabilizing negative curvature.

Future work should move beyond rolling ball bearings to systems that better approximate point-like dynamics. One promising approach is the use of Leidenfrost levitated LN\textsubscript{2} beads, which exhibit significantly reduced friction. Preliminary experiments with such systems by our team have already demonstrated their viability and offer the potential for more precise trajectory comparisons with theoretical models, deepening our understanding of asymmetric saddle dynamics and their broader implications, from ion traps to astrophysical systems.

\begin{acknowledgments}

 We are grateful for helpful discussions with Daniel Davis, Gregory Kestin and Logan McCarty. We gratefully acknowledge support from the Faculty of Arts and Sciences at Harvard University. GPT-4 \citep{OpenAI2023ChatGPT4} was used to refine this manuscript. 

\end{acknowledgments}




\nocite{*}
\bibliography{references} 

\end{document}